\documentstyle[aps,preprint]{revtex}
\newcommand{\be}{\begin{equation}}
\newcommand{\ee}{\end{equation}}
\newcommand{\bea}{\begin{eqnarray}}
\newcommand{\eea}{\end{eqnarray}}
\newcommand{\nn}{\nonumber \\}
\begin{document}

\draft
\preprint{Alberta-Thy-4-94}
\title{
Exact solution for scalar field collapse}
\author{ Viqar Husain, Erik A. Martinez, Dar\'\i o N\'u\~nez
\footnote{Instituto de Ciencias Nucleares, UNAM,
70-543, M\'exico, D. F. 04510,  M\'exico. \\E-mails:
 husain@einstein.phys.ualberta.ca, martinez@phys.ualberta.ca,
 \hfill\break nunez@phys.ualberta.ca.} }
\address{Theoretical Physics Institute,\\
    University of Alberta,\\
  Edmonton, Alberta T6G 2J1, Canada.}
\maketitle
 \begin{abstract}

We give an exact spherically symmetric solution for
the Einstein-scalar field system. The solution may be interpreted
as an inhomogeneous dynamical scalar field cosmology. The spacetime has
a timelike conformal Killing vector field and is asymptotically conformally
flat. It also has black or white hole-like regions containing trapped
surfaces. We describe the properties of the apparent horizon  and
comment on the relevance of the solution to the recently discovered
critical behaviour in scalar field collapse.

\end{abstract}
\vspace{7mm}
\vfill
\eject

The gravitational collapse of distributions of matter is one of
the most important research areas in general relativity. The essential
question posed is whether and under what initial conditions a black hole
or naked singularity forms in the collapse. One of the aims of
studying such problems is to test the cosmic censorship conjecture
\cite{pen}, one form of which states that gravitational collapse
produces black holes.

A model system for studying this question is provided by the Einstein
equations minimally coupled to a massless scalar field.
While this full system appears to be  intractable, the simplified set of
equations obtained by imposing spherical symmetry is easier
to handle. Without any
matter fields, the spherically symmetric metric does not contain any
field degrees of freedom. Therefore, with a scalar field, the system is
effectively a two dimensional field theory and it can be described
by a single two dimensional nonlinear differential equation

\cite{chris,don}.

There are a number of exact solutions known for this system, almost all of
which are either static or depend only upon the time coordinate
\cite{jnw,ssol}. The first non-static solutions has been given by Roberts
\cite{roberts}, (which are different from the one we give below).
The equations have been studied in detail by Christodoulou \cite{chris} who
established, among other things, that there exist regular solutions for
arbitrarily long times for particular types of initial data.

The model has also been studied numerically and there are a number
of interesting numerical results.  The first results obtained by
Goldwirth and Piran \cite{dal} indicated
that there is a class of initial data that leads to black hole
formation. More recently it has been shown by Choptuik \cite{chop}
that, for large classes of initial data,  there is critical behaviour
at the onset of black hole formation: the black hole mass $M_{BH}$ is
given by the equation $M_{BH} = K |c-c_*|^\gamma$, where $K$ is a constant,
$c$ is any one of the  parameters in the initial data for the scalar field,
$c_*$ is a critical value of the parameter, and $\gamma\sim.37$ is a
universal exponent. The remarkable feature of this result  is that $\gamma$
appears to be independent of a set of particular shapes of the initial data,
and is universal in this sense. It has been shown by Abraham and Evans
\cite{ae} that the same mass equation is obtained for the axisymmetric
collapse of gravitational radiation. Thus the critical behaviour appears
to be independent of not only the type of matter fields, but also the
symmetries of the system.

It would be very useful to understand the universality of this result
analytically.  A modest approach is to attempt to find an exact solution
describing scalar field collapse and to see if one can read off the
critical  behaviour by calculating the mass of the black hole.

Here we describe an exact solution for scalar field collapse and
discuss some of its properties. While the solution we present does
not describe a realistic collapse corresponding to the classes of initial
data used in the numerical work mentioned above,  it appears to be among
the few exact non-static solutions known for this system.

The Einstein-scalar field equations we consider (in units $G=c=1$ )
are
\be
G_{\mu\nu} = 8\pi T_{\mu\nu}; \ \ \ \ \ \ \
T_{\mu\nu} = \phi_\mu\phi_\nu
- {1\over 2} g_{\mu\nu} g^{\alpha\beta}\phi_\alpha\phi_\beta.
\ee
which may be written in the form
\be
R_{\mu\nu} =  8\pi \phi_\mu\phi_\nu,\label{eq:ric}
\ee
(where the subscript on $\phi$ denotes partial differentiation).

The spherically symmetric solution we obtain is
\be
ds^2=(at+b)\,(-f^2(r)dt^2 + f^{-2}(r)dr^2) + R^2(r,t)
(d\theta^2 +\sin ^2\theta d\phi^2), \label{metric}
\ee
where
\bea
f^2(r) &=& ( 1 -  2 c / r)^\alpha  \nn
R^2(r,t) &=& (at+b) r^2  (1- 2 c/ r)^{1 - \alpha}.
\eea
 The scalar field is
\be
\phi(r,t) = \pm {1\over 4\,\sqrt{\pi}} \,
\ln[d\,( 1- 2 c /r )^{\alpha\over \sqrt{3}}\,(at+b)^{\sqrt{3}}],
\label{scal}
\ee
where $a,b,c,d $ are  constants,

$\alpha = \pm \sqrt{3}/2$, and the overall sign of $\phi$ is independent

of $\alpha$. The
coordinate ranges and the values of the constants are
coupled (for the metric to be Lorentzian):
  $-b/a \le t \le \infty$ and $ 2c\le r\le \infty$ ($c>0$).

We note that for $a\ne 0$ there is a coordinate transformation, that
eliminates the constant $b$. However, since we will also be interested
in the metrics for which $a=0$, it is useful to keep the
general form (\ref{metric}). We note also that when $b=0$, there
are two ranges for $t$, namely $ 0 \le t \le \infty$ for $a>0$,
and $ -\infty \le t \le 0$ for $a<0$. As shown below these correspond
to white and black hole-like solutions respectively, and simply reflect
the choice of the arrow of time.

When $a\ne 0$, the coordinate transformation $T=at+b$ does not eliminate

$a$ in the metric, which is an ovserall scale. The constant $d$ in the

scalar field is a trivial additive constant.

 The only Killing vectors of the metric (3) are the three
 associated with spherical symmetry.
The  metric also has the conformal Killing vector field
$V=\partial/\partial t$ such that
\be
{\cal L}_V g_{\mu\nu}= {a\over at+b} g_{\mu\nu}.
\ee
Asymptotically ($r\rightarrow\infty$), the metric is conformal to the
Minkowski metric. The locus of points at $r=2 $  is a timelike curvature

singularity, and so the `horizon' is shrunk to
a point as in the static metric given by Janis, Newman and Winicour (JNW),
and others \cite{jnw,ssol}.
There is also a spacelike singularity at $t=-b/a$. The
solution may be interpreted as a scalar field cosmology, since it is
not asymptotically flat.

 For the JNW metric, the functions $f(r)$
have arbitrary exponents specified by integration constants rather than
the fixed  $\alpha = \pm\sqrt{3}/2$ above. Thus our metric is conformal to
the JNW metric  with this exponent fixed, and  with $(at+b)$ as the
conformal factor. In fact for $a=0$ we recover one of the JNW metrics.
Therefore  $a$  distinguishes  static from

non-static metrics. This is similar to the parameter $k$ in

Friedman-Robertson-Walker cosmologies, which discretely distinguishes

the spatial curvatures of the metrics.

The case $c=0$ gives a homogeneous time dependent solution. This
is equivalent to the $r\rightarrow \infty$ limit of (\ref{metric}).
The parameter $c$ therefore distinguishes  homogeneous from inhomogeneus
time dependent solutions.

For comparison with recent numerical work \cite{chop,ae}, where the scalar

field and its time derivative are specified as part of the initial data on

a spacelike hypersurface, we note that this data for our solution is

\bea
\phi (r,t=t_0) &=&  {1\over 4\,\sqrt{\pi}}\,
\ln(d\,(1- 2\,c/r )^{{\alpha\over {\sqrt{3}}}}\,
(at_0 + d)^{\pm\sqrt{3}}),\nonumber \\
\dot{\phi} (r, t=t_0) &=&  {1\over {4(at_0 + b)}}\,\sqrt{{3\over\pi}}.
\label{id}
\eea
The asymptotic behaviour of $\phi$ is
\be
\phi (r = \infty, t=t_0) =  {1\over 4\,\sqrt{\pi} }\,
\ln(d \,(at_0 + b)^{\pm\sqrt{3}} ).
\ee
These are not the initial data associated with the standard collapse
situation  \cite{chop,ae}, where the data is typically an ingoing pulse

with a specified amplitude and width.

The Ricci scalar is
\be
{\cal{R}} = {12 c a^2 (r-c) - 3 a^2 r^2 \over 2    r^2 (at+b)^3}
(1-{2c\over r})^{ -2 - \alpha}
 + {2   c^2 (1-\alpha^2) \over (at+b) r^4} (1-{2c\over r})^{-2 + \alpha }
 \ee
which  shows that  curvature singularities are present at $r=2c$, and at
  $t=-b/a$.

Since the metric is not static, it is
of interest to investigate the existence and properties of the
apparent horizon and to calculate the mass function.
The apparent horizon is the 3-surface on which outgoing or ingoing  null
rays are momentarily stationary. The presence of the horizon for
asymptotically flat spacetimes indicates that there is a black or
white hole, and a measure of the mass within it is given by the mass
function evaluated at the horizon. This measure of the black hole
mass has been used in recent numerical work on the collapse of ingoing
matter pulses \cite{chop,ae}. Although the  metric (\ref{metric}) is not
asymptotically flat, we can nevertheless see whether there are horizons.

The apparent horizon surface is given
by
\be
g^{\alpha\beta} R,_{\alpha} R,_{\beta} = 0,
\ee
which for our metric gives the equation
\be
 {a \over at_{AH} +  b} = { 2  \over r^2}[r-c\,(1 + \alpha)]
(1- 2\,c/ r  )^{ \alpha - 1 }
\label{ah}
\ee
This equation has no non-trivial solution for $a=0$ which corresponds to
the static JNW metric, wheras for  $a\ne 0$ there is always an
evolving apparent horizon.

The apparent horizon can in general be spacelike, null or timelike
in different spacetime regions. This is easily determined by
calculating the ratio of the slopes of the apparent horizon
 and the outgoing null ray. This ratio for our metric (with $a\ne 0$) is
\be
 { t_{AH,r} \over t_{N,r} } =
  1 - { (1 - 2c/r)\over 2[ 1 - c (1+\alpha)/r]^2}.
\label{slope}
\ee
  The second term on the right hand side
is always positive, therefore the apparent horizon is spacelike
for all $r > 2c$. It is null only at $r=2c$.

   The scalar field is not singular at the apparent horizon.

Figures 1 and 2 are plots of the horizon for $\alpha=\pm \sqrt{3}/2$
 and $ c=1, b=0$. The  main features remain unaltered for all values of
 $b,c$. The horizon forms at $r=2c,\ t=0$ and grows in size
forever. For $\alpha = \sqrt{3}/2$, the light cones are collapsed to
a vertical line at $r=2c$ and open up to a slope of $\pm 1/{b^2}$ as
$r\rightarrow \infty$, whereas for $\alpha = -\sqrt{3}/2$ the cones
collapse to a horizontal line at $r=2c$.

It is of interest to note a number of other features of the
apparent horizon. By computing the expansions of the spacelike
symmetry 2-spheres
\be
 ds^2 = R^2(r,t)(d\theta^2 +{\sin}^2\theta d\phi^2)
\ee
along future pointing null directions orthogonal to the
spheres, one can determine whether the apparent horizon is past
or future, and inner or outer, and what region is trapped (see
\cite{hay} for a general discussion of apparent horizons).

 The expansions $\theta_{\pm}$ of the area 2-form
$\omega = R^2 (r,t)\sin\theta\ d\theta\wedge d\phi$
of the  2-spheres are defined by
\be
{\cal L}_{l_\pm}\omega = \theta_{\pm}\omega,
\ee
where ${\cal L}$  denotes the Lie derivative and
\be
 l_+ := {\partial\over \partial \, t}+f^2\,{\partial\over \partial \, r}
 \,\, \ \ \
 l_- := {\partial\over \partial \, t}-f^2\,{\partial\over \partial \, r}
\ee
are the outgoing and ingoing future pointing null directions.
With $b=0$ and $a>0$, (so that $0\le t\le \infty $), we find
\be
\theta_\pm  = {1\over t} \pm  {1\over t_{AH}},
 \ee
 with $t_{AH}$ as given in Eq. (11).
Thus it is  the ingoing expansion $\theta_{-}$ that vanishes at the
apparent horizon, while the outgoing expansion is $\theta_+ = 2/t_{AH}>0$.
This implies that the horizon is a past horizon. For a given value of $r$,
the symmetry 2-spheres are trapped surfaces for $t < t_{AH}$, since this
is the region where both the ingoing and outgoing light rays have positive
expansions.  Similarly, the region $t>t_{AH}$ is a normal region where the
outgoing light expansion ($\theta_+$) is positive and the ingoing one
($\theta_-$) is negative.

We  note also that if ${\cal L}_{l_+}\theta_- |_{AH}<0$ the
horizon is an outer one (otherwise it is inner). For the
metric (\ref{metric}), we find
\be
{\cal L}_{l_+}\theta_- |_{AH} = {1 \over t_{AH}^2}
[{ t_{AH,r} \over t_{N,r} } - 1 ].
\ee
{}From (\ref{slope}) it follows that this is always less than zero
except  at the singularity where it is zero.

Summarizing the above results, the apparent horizon is a past outer
one and is spatial everywhere except at $r=2c$, where it is
null. The scalar field flows from the past trapped region $t < t_{AH}$
 (white hole) into the untrapped region $t>t_{AH}$. An observer in
 the untrapped region sees both the $t=0$  initial singularity and the one
 at $r=2c$.

  When $b=0$, there is a change for the
time reversed case which corresponds to $a<0$ and $-\infty\le t \le 0$.
The horizon is now a future outer one and corresponds to a black hole
situation. The future singularity at $t=0$ is covered by the
spacelike horizon and the region is a black hole.

We now turn to a discussion of the mass function defined by
\be
m(r,t) = {R\over 2}( 1 - g^{\alpha\beta} R,_{\alpha} R,_{\beta}).
\label{mf}\ee
This function for static or stationary asymptotically flat spacetimes
gives the ADM mass in the asymptotic limit.  Its value on the apparent
horizon is obtained by substituting
$t_{AH}(r)$ from (\ref{ah}) into (\ref{mf}):
\be
M:= {R_{AH} \over 2} = m(r,t)|_{AH} = \sqrt{{|a| \over 8}}
{ r^2\,(1- 2\,c/ r )^{1-\alpha} \over
  \sqrt{ r-c\,( 1 + \alpha )} }.
\label{mass}
\ee
The mass $M$ is zero at $r=2c$ and grows as $r^{3/2}$ for large $r$.
We note that it is independent of $b$. For $a=0$, $M=0$, which
corresponds to the static JNW solution. This corresponds to the
fact that there is no solution to the apparent horizon equation
(\ref{ah}) for $a=0$.

Superficially, the mass function may be viewed as describing a form
of critical behaviour since for fixed $r$ we can write
$M=constant(a-a_*)^{1/2}$, with the `critical' value $a_* = 0$, and with
one half as the `exponent'. Similarly, one can do this for the
parameter $c$ in (\ref{mass}), which gives a different exponent.
This, however, does not shed light on the collapse
of initial pulses of scalar field, since the initial data (\ref{id})
does not correspond to this situation.

The metric is nevertheless an exact solution and may  be
viewed as a model in which the formation and evolution of
an apparent horizon can be studied exactly. The value of the
parameter $a$ in the solution determines whether an apparent horizon
forms. It would be of much interest to find an exact solution
corresponding to a realistic collapse, as this would provide an
analytical model for viewing the scaling and critical behaviour
obtained in recent numerical work \cite{chop,ae}.

We would like to thank Werner Israel, Don Page, Patrick Brady, and
especially Henrique P. de Oliveira for stimulating discussions.
V. H. and E. M. were  supported by the Natural Science and Engineering
Research Council of Canada. D. N. thanks Direccion General de Asuntos
del Personal Academico, UNAM, for partial support.

\vfill\eject
\centerline {\bf FIGURE CAPTIONS}
\smallskip
\parindent=0pt \parshape=3 0truein 6truein .7truein 5.3truein .7truein
5.3truein
Figure 1:~The apparent horizon for $\alpha = +{\sqrt{3} \over 2}$ and $
c=1, b=0$. The light cones illustrate the spacelike character of the horizon.
The singularities are at $r=2c$ and $t=0$. The trapped region is
$t<t_{AH}$.

\smallskip
\parindent=0pt \parshape=3 0truein 6truein .7truein 5.3truein .7truein
5.3truein
Figure 2:~The apparent horizon for $\alpha = -{\sqrt{3} \over 2}$ and $
c=1, b=0$.
\vfill\eject


\begin{references}

\bibitem{pen} R. Penrose, Riv. Nuovo Cimento {\bf 1}, 252 (1969).

\bibitem{chris} D. Christodoulou, Comm. Math. Phys. {\bf 105}, 337 (1986);
  {\bf 106}, 587 (1986); {\bf 109}, 591 (1987); {\bf 109}, 613 (1987).

\bibitem{don} D. N. Page, private communication.


\bibitem{jnw} A. I. Janis, E. T. Newman, J. Winicour,  Phys. Rev. Lett.,
{\bf 20}, 878, (1968).

\bibitem{ssol} O. Bergman, R. Leipnik, Phys. Rev. {\bf 107}, 1157 (1957);
H. A. Buchdahl, Phys. Rev. {\bf 111}, 1417 (1959);
M. Wyman, Phys. Rev D {\bf 24}, 839 (1981);
A. Agnese, M. LaCamera, Phys. Rev. D {\bf 31}, 1280 (1985);
S. Abe, Phys Rev. D {\bf 38}, 1053 (1988); T. Papacostas,
Jour. Math. Phys. {\bf 32}, 2468 (1991).

\bibitem{roberts} M. D. Roberts, Gen. Rel.and Grav. {\bf 21}, 907 (1989).


\bibitem{dal} D. S. Goldwirth and T. Piran, Phys. Rev. D {\bf 36},
3575 (1987).

\bibitem{chop} M. Choptuik,  Phys. Rev. Lett. {\bf 70}, 9 (1993).

\bibitem{ae} A. Abraham and C. R. Evans, Phys. Rev. Lett.
{\bf 70}, 2980 (1993).


\bibitem{hay} S. Hayward, `General laws of black-hole dynamics',
 preprint Max Planck Institute (March 1993).

\end{references}
\end{document}